\documentclass[twocolumn,showpacs,aps,prl,superscriptaddress]{revtex4-1}
\usepackage[utf8x]{inputenc}
\usepackage{graphicx}
\usepackage{epsf}
\usepackage{natbib}
\begin{document}

\title{Silicene vs. 2D ordered silicide: the atomic and electronic structure of the Si-$(\sqrt{19}\times\sqrt{19})R23.4^{\circ}$/Pt(111)}

\author{M. \v{S}vec}
\email[corresponding author: ]{svec@fzu.cz}
\author{P. Hapala}
\affiliation{Institute of Physics AS CR, Cukrovarnick\'a 10, CZ-16200 Prague, Czech Republic}
\author{M. Ondr\'{a}\v{c}ek}
\affiliation{Institute of Physics AS CR, Cukrovarnick\'a 10, CZ-16200 Prague, Czech Republic}
\author{P. Merino}
\affiliation{Centro de Astrobiolog\'ia INTA-CSIC, Ctra. de Ajalvir, km.4, ES-28850 Madrid, Spain}
\author{M. Blanco-Rey}
\affiliation{Dpto. de F\'{\i}sica de Materiales UPV/EHU, Apartado 1072, 20018 Donostia-San Sebasti\'an, Spain}
\affiliation{Donostia International Physics Center, P$^o$ Manuel de Lardizabal 4, 20018 Donostia-San Sebasti\'an, Spain}
\author{P. Mutombo}
\affiliation{Institute of Physics AS CR, Cukrovarnick\'a 10, CZ-16200 Prague, Czech Republic}
\author{M. Vondr\'a\v{c}ek}
\affiliation{Institute of Physics AS CR, Cukrovarnick\'a 10, CZ-16200 Prague, Czech Republic}
\author{Y. Polyak}
\affiliation{Institute of Physics AS CR, Cukrovarnick\'a 10, CZ-16200 Prague, Czech Republic}
\author{V. Ch\'{a}b}
\affiliation{Institute of Physics AS CR, Cukrovarnick\'a 10, CZ-16200 Prague, Czech Republic}
\author{J. A. Mart\'{i}n Gago}
\affiliation{CSIC-ICMM, C/Sor Juana Ines de la Cruz 3, E-28049 Madrid, Spain}
\author{P. Jel\'{i}nek}
\affiliation{Institute of Physics AS CR, Cukrovarnick\'a 10, CZ-16200 Prague, Czech Republic}
\affiliation{Graduate School of Engineering, Osaka University 2-1, Yamada-Oka, Suita, Osaka 565-0871, Japan}


\begin{abstract}
We discuss the possibility of a 2D ordered structure formed upon deposition of Si on metal surfaces. 
We investigate the atomic and electronic structure of the Si-$(\sqrt{19}\times\sqrt{19})R23.4^{\circ}$/Pt(111) surface reconstruction by means of a set of experimental surface-science techniques supported by theoretical calculations. The theory achieves a very good agreement with the experimental results and corroborate beyond any doubt that this phase is a surface alloy consisting of Si$_3$Pt tetramers that resembles a twisted Kagome lattice. These findings render unlikely any formation of silicene or germanene on Pt(111) and other transition metal surfaces.
\end{abstract}

\maketitle

The experimental discovery of graphene \cite{Graphene_Novoselov04} has stimulated a~search for other families of two-dimensional crystals \cite{Butler_ACSNano13, Geim_Nature13}.  Probably the most desired 2D material for nanoelectronic industry is the silicon-based counterpart of graphene, the so-called silicene. This request has initiated an~intensive debate about its existence. Recently several groups reported formation on the Si overlayers with different periodicities, when Si is deposited on Ag(111) (e.g. \cite{Kara_SSP12,Vogt_PRL12}). Some of these are claimed to be silicene, but its sp$^2$ electronic configuration is widely disputed \cite{Cahangirov_PRL09,Lalmi_APL10,Lin_PRL13,Vogt_PRL12,Majzik_JPCM13}. Recently Meng et. al. \cite{Meng_NanoLett13} observed the formation of ($\sqrt{7}\times\sqrt{7})$R19.1$^o$ ($\sqrt{7}$ in the following) Si superstructure on the Ir(111) surface. They assigned it to a~strongly buckled silicene sheet formed on top of the Ir(111), based on a~rough agreement between Scanning Tunneling Microscopy (STM), low-energy electron diffraction (LEED) and density functional theory (DFT) calculations.

On the other hand, silicon is well known to form binary alloys with the majority of the transition metals \cite{phase_diagrams}. Thus, when Si is adsorbed onto a pure transition metal, it dissolves in the bulk upon annealing. However, by tuning the metal temperature, Si can also segregate to the surface. The resulting structures vary from surface alloys to overlayers and exhibit interesting atomic arrangements. For example, Si adsorbed on Cu(110) substrates at room temperature intermixes and forms a c(2$\times$2) surface alloy superstructure \cite{Gago_PRB97}.  Earlier studies of the segregation of Si impurities on the surface of Pt(111) found the ($\sqrt{19}\times\sqrt{19}$)R23.4$^o$ superstructure ($\sqrt{19}$) and also a $\sqrt{7}$\cite{diebold,Nashner_JPCHB102}. These structures have been determined to be a silicide using spectroscopic methods. However their precise atomic structure has not been resolved yet due to a~lack of adequate techniques and/or limited computer resources. Therefore, the missing atomic structure and the advent of silicene \cite{Kara_SSP12,Vogt_PRL12} have revived the intensive research of such systems. 

The surface structure of a Si-covered metal is very sensitive to the strength of the mutual interaction between the Si atoms and the metal substrate, thus the structures may vary significantly as a function of the atomic species involved. Nevertheless, similarities exist among the above mentioned Si/Pt(111) superstructures -- $\sqrt{19}$, $\sqrt{7}$ -- and the analogous systems: Ge/Pt(111) and Si/Ir(111) \cite{Ho_SurfSci08, Meng_NanoLett13}.  These superstructures have been studied, up to now,  using separately either the LEED patterns, STM images, Auger or photoelectron spectroscopy.  At a first glance, these structures resemble the Moir\'e patterns reported for graphene on metals, and particularly on Pt(111) \cite{merino2011}.
However, the lack of atomic resolution in the STM images and the lack of quantitative analyses of LEED data account for the missing precise atomic determination of this type of systems.  We want to stress that, to our knowledge, a comprehensive study combining several complementary experimental and theoretical techniques, which would rule out/confirm the existence of silicene on transition metals, has been missing so far.

In this letter, we characterise the atomic and electronic structure of the $\sqrt{19}$ surface reconstruction of the Si/Pt(111) and present strong evidence that it is a 2D surface alloy. The ultimate aims of this work are: (i) to decide unambiguously whether Si can form a 2D honeycomb lattice on Pt(111) surface -- the silicene; (ii) to determine precise atomic structure of the $\sqrt{19}$; and (iii) to find the extent of hybridization between the Si and Pt orbitals.
For this purpose, we use an integral set of experimental and theoretical tools.

\begin{figure*}
\includegraphics{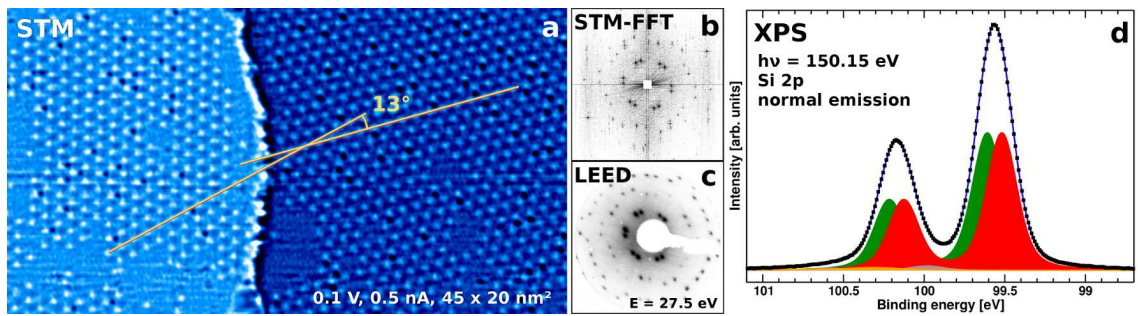}
\caption{(Color online) Mesoscopic characterization of the Si-$(\sqrt{19}\times\sqrt{19})-R23.4^{\circ}$/Pt(111) surface structure. (a) STM image of two terraces divided by a step, each hosting one of the two domain orientations of the structure. Yellow lines denote the mutual orientation between the domains. (b) and (c) show the comparison of a power spectrum obtained from the STM image by FFT and a LEED pattern obtained at electron energy of 27.5~eV, respectively. (d) photoemission doublet Si-$2p$ fitted with two components.}
\label{fig1}
\end{figure*}

Based on the experimental results, we deduce a candidate for the Si-$\sqrt{19}$/Pt(111) atomic structure and confirm it using Density Functional Theory (DFT) calculations. Here, we employ a~collection of different experimental techniques, operating in ultrahigh vacuum conditions, including STM, dynamic atomic force microscopy (dAFM) with a tuning fork \cite{Berger_JBN13}, quantitative full-dynamic analysis of intensity-energy curves in LEED (LEED-IV), synchrotron radiation photoemission spectroscopy (SRPES) and angle-resolved ultraviolet photoemission (ARUPS) measurements. The experimental results are supported by the total energy DFT and simulations of STM images, ARUPS patterns and LEED-IV curves. A detailed description of the experimental and the computational methods can be found in the Supplemental information \cite{supp}.

The sample preparation consisted of cleaning the Pt(111) surface by cycles of Ar$^{+}$ sputtering and annealing to 1200~K. The surface was exposed to $1\times 10^{-7}$ O$_2$ for 30~sec during the last annealing cycle, in order to remove any C residue from the surface. Si was evaporated from a wafer stripe ($\approx 10\times 5\times 0.5$~mm$^3$), by direct current heating to 1400~K, which produced a flux of $\approx$0.06~ML~/~min. Si was deposited on the sample held at 750K. This preparation procedure produced the $\sqrt{19}$ reconstructed surface, manifested by a sharp LEED pattern.

Fig.~\ref{fig1}a shows a~STM image of the $\sqrt{19}$ phase covering over 80~\% of the Pt(111) surface. The imaged area contains two chiral domains, which have an angular offset of $\approx$13$^{\circ}$ (ideal value 13.17$^{\circ}$). In this particular image, the domains are separated by a step edge. However, both domains have also been observed coexisting within the same terrace. The fast Fourier transform (FFT) power spectrum in Fig.\ref{fig1}b of the STM image is almost identical to the LEED pattern obtained at 27.5~eV, shown in Fig.\ref{fig1}c, which contains spots corresponding to both domain orientations.

\begin{figure}
\includegraphics{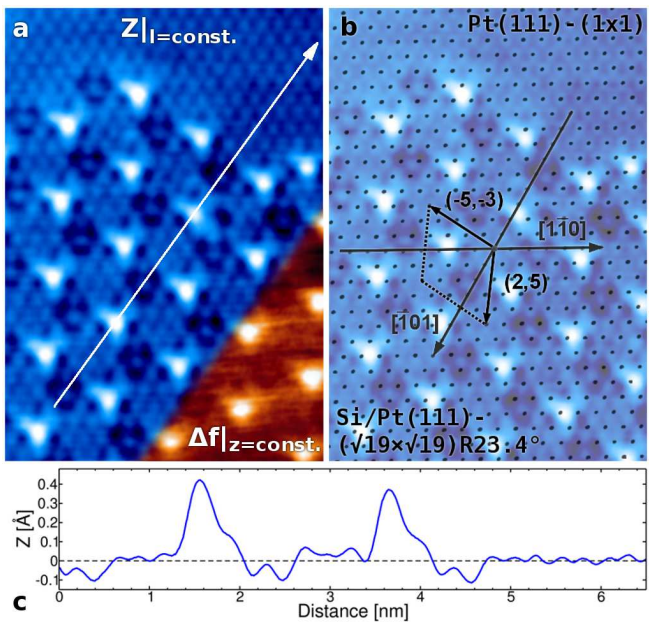}
\caption{
(Color online) (a) Atomically-resolved STM topography of a Pt(111) region ($5\times7.2$~nm$^2$) mostly covered with Si-$(\sqrt{19}\times\sqrt{19})R-23.4^{\circ}$ structure, filled states, -50~mV, 7~nA. The triangular inset shows the $\Delta$f signal obtained by a simultaneous STM/dAFM measurement in the constant height and repulsive mode \cite{supp}. (b) The grid corresponding to the Pt atom positions on the ($1\times1$) lattice superimposed onto the STM image. Surface lattice vectors are determined from the orientation of the main surface crystallographic axes and the periodicity of the ($1\times1$). (c) topographic profile of the transition between the $\sqrt 19$ and the ($1\times1$) regions along the arrow in (a).
}
\label{fig2}
\end{figure}

Atomically resolved STM images taken at low bias and high tunneling current reveal a complex pattern within the $\sqrt{19}$ unit cells -- see Fig.~\ref{fig2}a. Essentially, parts of the $\sqrt{19}$ superstructure seem to be a continuation of the Pt(111)-($1\times1$) surface structure, modified by triangular-shaped inclusions of dark and bright contrast. The line profiles plotted on Fig.~\ref{fig2}c reveal that both the $\sqrt{19}$ phase and the clean Pt(111) surface are located in the same atomic layer. These observations make it unlikely that the $\sqrt{19}$ phase consists of an extra atomic Si layer -- e.g. silicene -- formed on top of the clean Pt surface. By overlaying a hexagonal grid corresponding to the ($1\times1$) periodicity and extrapolating it to the region of the $\sqrt{19}$ superperiodicity (Fig.\ref{fig2}b), the unit cell vectors $(2,5)$ and $(-5,3)$ of the superstructure can be identified with respect to the base vectors defining the ($1\times1$) lattice (based on the directions along the $[\bar{1}01]$ and $[\bar{1}\bar{1}0]$ crystallographic axes of the surface).

One bright and one dark triangular-shaped features are found within each unit cell marked in Fig.~\ref{fig2}b. Both are located at the centers of the unit cell halves, which coincide with the threefold symmetry axes of the $\sqrt{19}$ reconstruction. Both the bright and dark triangles have their maximum intensity at the center. In a direct interpretation, this could be assigned to atoms at interstitial positions of the ($1\times1$) surface lattice. Moreover, the dAFM image (inset of the Fig.\ref{fig2}a) taken in the constant height mode and the repulsive mode shows protrusions in the centers of the bright triangles, indicating that the corresponding interstitial Pt atom lies certainly higher than the rest.

We performed high-resolution SRPES with an excitation energy of 150~eV in the normal emission geometry. 
The measurements revealed a Si-2p doublet located at 99.56~eV (binding energy) that emerges
after deposition of Si -- see Fig.~\ref{fig1}d. The energy position of the doublet corresponds to a typical silicide \cite{silicide_BE}. The small width of the peaks ($\approx$0.3~eV) and the apparent lack of a fine structure hint that all the Si atoms in the $\sqrt{19}$ structure are in a~similar chemical state. The deposition of Si has also influence on the Pt-$4f$ component representing the substrate, which loses a characteristic surface-related feature (shown in \cite{supp}). From an intensity evaluation of the Si-$2p$ and Pt-$4f$ peaks, the estimated coverage of Si atoms is $\approx$0.3~ML. 

\begin{figure}
\includegraphics{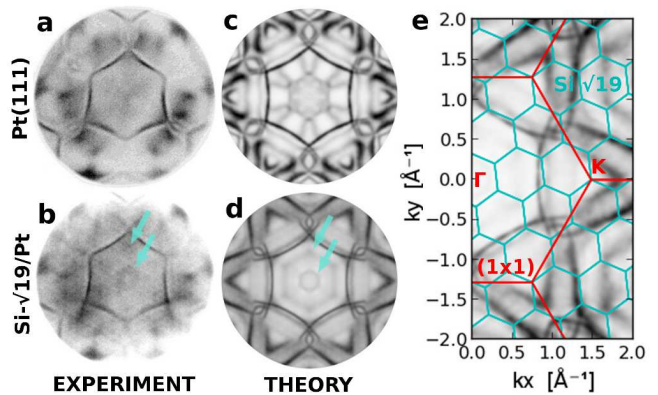}
\caption{(Color online) (a), (b) ARUPS $k$-space maps of clean Pt(111) and Si-$(\sqrt{19}\times\sqrt{19})R23.4^{\circ}$/Pt(111) at the Fermi level, (c), (d) DFT-calculated band maps for the best-fit model and projected to higher Brillouin zones. The arrows mark the emerging features due to formation of the structure. (e) Map of the first two BZs of the (1$\times1$) (red) and the $\sqrt{19}$ (cyan) structures. 
}
\label{fig4}
\end{figure}

Furthermore, we mapped the electronic structure in the $k$-space near to the the Fermi level using ARUPS.  Fig.~\ref{fig4}a shows the clean Pt(111)-($1\times1$) band structure with two characteristic features: (i) a~hexagonal shape rising from a~parabolic $sp$-like band when crossing the Fermi level; and (ii) weak features spreading from the center of the Brillouin zone (BZ). The band map of the $\sqrt{19}$ phase maintains the main features, as observed in Fig.~\ref{fig4}b. However, there is a considerable suppression of the overall intensity and a~visible change around the center of the BZ (marked by arrows). A shape surrounding the $\Gamma$ point is discernible and intensity also grows near to the main band vertices. 

In the following, we will uncover a~suitable atomistic model for the $\sqrt{19}$ structure that fulfills all the experimental findings. The SRPES data implies a very similar chemical environment for all Si atoms as well as a strong hybridization with its Pt host. The STM profiles in Fig.\ref{fig2}c taken across the $\sqrt{19}$ and ($1\times1$) regions of the sample shows a negligible height difference, indicating that Si atoms are most likely {\it embedded} in the Pt(111) surface. Most of the protrusions in the STM image can be assigned to the regular (1$\times1$) lattice extrapolated from the clean Pt(111), except for the protrusions centered in the dark and bright triangular features.

We constructed a~set of atomistic models derived from the clean Pt(111) $\sqrt{19}$ supercell. In these models, the three atoms nearest to the center of both half-unit cells were replaced by one Si, one Pt atom or 4-atom pyramids (tetramers). The tetramers consisted of a combination of Si and Pt atoms. The individual models are described in the Supplemental information \cite{supp}. In addition, we also considered a~silicene model consisting of a ($3\times3$) Si honeycomb lattice aligned on top of the Pt $(1\times1)$ surface resulting in the $\sqrt{19}$ periodicity. The models were subjected to a geometry optimization using total energy DFT calculations with the Fireball code \cite{Review-Fireball2011}. Subsequently, we calculated STM images using the optimized structure of each model. The two most thermodynamically stable models and the silicene model, were further optimized by the plane wave code VASP \cite{kresse1996a}, but the relative stabilities of the studied models were not altered \cite{supp}. 

\begin{figure}
\includegraphics{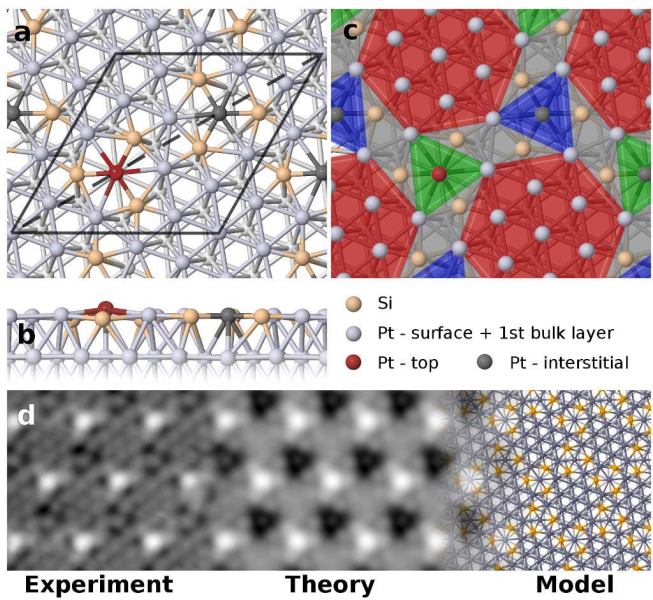}
\caption{(Color online) (a) and (b) Top and side view of first two layers of the best-fit model structure. 
(c) A scheme of the twisted Kagome structure adopted by the system in the presence of the Si$_3$Pt tetramers. The overlay represents lateral positions of Si, Pt atoms in the topmost layer. (d) A blend of a high-resolution STM topography, a simulated STM image -- both at -20mV -- and the model.}
\label{fig3}
\end{figure}

The atomic structure of the calculated thermodynamically most stable model is presented in Fig.\ref{fig3}a,b. The structure consists of two inserted tetramers formed by three Si atoms at the base 
and one Pt atom on top (Si$_3$Pt). The structure optimization results in a strong relaxation of the inserted tetramers. The top Pt atoms are pushed into the surface and the Si atoms are displaced outwards in the surface plane. The vertical relaxation of both Pt atoms of the tetramers depends strongly on the registry with the first subsurface Pt layer. The Pt adatom that is placed above a bulk Pt atom (red color) remains higher than the one above a triangle of bulk Pt atoms (dark gray). The calculated total charge transfer shows a strong charge depletion of the Si atom region. 

The calculated relative shift between the Si-$2p$ core levels of the Si atom triplets in our model is 52~meV, which is comparable to the experimentally observed energy difference 90~meV for a two-component fit of the photoemission Si-2p doublet in Fig.1c.
Inclusion of the tetramers effectively divides the Pt surface layer into basic building blocks, as shown in Fig.\ref{fig3}c; hexagonal areas (red), which essentially preserve the (1$\times$1) structure, and triangular regions (blue, green) arising from the two types of tetramers. In this simplified scheme the system resembles a twisted Kagome lattice \cite{kagome}.

We note that not only the total energies but also the calculated STM images, LEED-IV curves and ARUPS Fermi level bandmaps of this model provide the best agreement with the experimental evidence. The simulated STM image (Fig.\ref{fig3}d) follows very closely the charge density distribution. The agreement of the theory with the experiment is very good, since all experimentally observed features are reproduced. The bright spots corresponds to Pt atoms in the top position, the low current areas correspond to the positions of the Si atoms. The experimental images have a slightly better resolution than the theoretical simulation; this can be understood in terms of tip interaction with the surface atoms, which is known to enhance the positive atomic contrast on metals \cite{Blanco_PRB04}. 
Moreover, the dAFM images (Fig.\ref{fig2}a, \cite{supp}) taken in repulsive mode confirm the inequivalent height of the two tetramers within the $\sqrt{19}$. The enhanced repulsive interaction was detected only above a~location that exactly corresponds to the highest Pt atom in the proposed model.
 
The LEED-IV curves, measured on the fractional spots and the integer beams of the $\sqrt{19}$, were used as input for the computational optimization of the two most favorable models and the silicene \cite{supp}. For the proposed model, the Pendry reliability factor \cite{pendry_factor} reached a~value of 0.33, which is an acceptable value considering the complexity and extent of the model structure. On the other hand, we did not achieve values better than 0.7 for the silicene and the other candidate model. 

We also projected the band structures of the proposed model and the clean Pt(111) surface to higher BZs, as shown in Fig.\ref{fig4}. The leading hexagonally-shaped $d$-band, characteristic of the Pt(111) surface, together with the features attached to its vertices are well reproduced, compared to the ARUPS data in both the clean and the Si/Pt alloy. The circular feature around the $\Gamma$ point of the BZ and the intensity maxima at the inner sides of the Pt band are also reproduced.

 In the next, we show that our model of the ordered tetramers can be adopted to other systems better than the silicene models.  Besides the $\sqrt{19}$ superstructure, the Si/Pt(111) surface can also occur in the $\sqrt{7}$ phase \cite{Nashner_JPCHB102}. We want to point out that the experimental characteristics of the Si/Pt-$\sqrt{7}$ are very similar to those of the Si/Ir-$\sqrt{7}$ structure \cite{Meng_NanoLett13}.  A~general atomistic model of the $\sqrt{7}$ can be constructed from our tetramer model easily. It consists of one Si$_3$Pt tetramer per each $\sqrt{7}$ unit cell, where the distance between the tetramers is very similar to their distance in the $\sqrt{19}$ structure. We performed total energy DFT calculations of both the Si$\sqrt{7}$/Pt(111) and Si $\sqrt{7}$/Ir(111) systems comparing our tetramer with the silicene model. Our calculations clearly indicate that the surface alloy model is thermodynamically more stable than the silicene model for both $\sqrt{7}$ phases (for details see \cite{supp}).

Apart from Si/Pt(111), the $\sqrt{19}$ superstructure has been also found in Ge/Pt(111) and Si-intercalated graphene/Ir(111) \cite{Ho_SurfSci08,grsiir}. Namely, the Ge/Pt(111) surface has a~very sharp $\sqrt{19}$ LEED pattern \cite{Ho_SurfSci08}. Since the Ge elementary properties are closely related to Si, it is very probable that the atomic structure of that surface is analogous to our model. To support this hypothesis, we performed again the total energy DFT calculations of the tetramer and the germaneness model on the Ge$\sqrt{19}$/Pt(111) phase  \cite{supp}. Again we found the tetramer model superior to the silicene on the Ge$\sqrt{19}$/Pt(111) phase. The arguments above suggest that the tetramer formation may be common to a broader range of systems featuring Si or Ge on transition metal (111) surfaces and render unlikely the formation of silicene.

In summary, we conducted the comprehensive study of 2D Si-based surface structures combining several complementary experimental and theoretical techniques, which has been missing so far.  All employed techniques point toward the same proposed atomistic model. We determined unambiguously the atomistic model of the Si-$\sqrt{19}$ being ordered 2D surface alloy instead of silicene. Our analysis rules out the formation of silicene or germanene structure on transition metal surfaces such as Pt(111) and Ir(111) and favors the formation of Si(Ge)$_3$Pt tetramers.

\acknowledgments{We acknowledge the support by GA\v{C}R, grant no.\ 14-02079S, M\v{S}MT project no. LM2011029 and the spanish project no. MAT2011-26534. P.M. acknowledges the R. C. Rod\'es grant. We are very grateful to E. Ortega, P. de Andr\'es, A. Walter, J. L. Checa and  F. M. Schiller for valuable discussions.}


%

\end{document}